\documentclass[aps,prl,preprint,superscriptaddress,nofootinbib,showkeys,floatfix]{revtex4}
\usepackage{amsmath,amssymb,graphicx}
\usepackage[dvips]{psfrag}
\begin{document}
\preprint{\parbox[b]{1in}{ \hbox{\tt PNUTP-09/A06} }}

\title{Holographic calculation of  hadronic   
contributions to muon $g-2$}


\author{Deog Ki Hong}
\email[]{dkhong@pusan.ac.kr}
\affiliation{Department of
Physics,   Pusan National University,
             Busan 609-735, Korea}
\author{Doyoun Kim}
\email[]{abistp00@phya.snu.ac.kr}
\affiliation{FPRD and Department of Physics and Astronomy,\\
 Seoul National University,  Seoul 151-747, Korea}             
\author{Shinya Matsuzaki}
\email[]{synya@pusan.ac.kr}
\affiliation{Department of
Physics,   Pusan National University,
             Busan 609-735, Korea}

\date{\today}

\begin{abstract}
Using the gauge/gravity duality, we compute the leading order hadronic (HLO) contribution 
to the anomalous magnetic moment of muon, $a_\mu^{\rm HLO}$. Holographic renormalization is used to obtain a finite vacuum polarization.  
We find  $a_\mu^{\rm HLO} 
=470.5 \times 10^{-10}$ in AdS/QCD with two light flavors, 
which is compared with the currently revised BABAR data 
estimated from $e^+ e^- \to \pi^+\pi^-$ events, 
$a_\mu^{\rm HLO}[\pi\pi]=(514. 1\pm 3.8) \times 10^{-10}$.
\end{abstract}


\maketitle

\section{Introduction}
\label{intro}
A recent high-precision measurement of $\pi^{+}\pi^{-}$ cross section  by the BABAR experiment~\cite{Aubert:2009fg} has found that the previous $e^{+}e^{-}$-based evaluation of the leading hadronic (HLO) contributions  to muon $g-2$  was fairly underestimated. With this new measurement the discrepancy between the $e^{+}e^{-}$ and $\tau$-based results for the dominant two-pion mode is narrowed to $1.5\,\sigma$ level in the dispersion integral~\cite{Davier:2009zi}. 
It is therefore quite necessary to find other means to test this new experimental data. 

To confirm the new BABAR data one needs to calculate it directly from QCD, for which 
Lattice QCD would  certainly be the best tool. However, the state-of-the-art calculations of lattice QCD still suffer from a large systematic error, compared to the experimental value, due to difficulties in simulating light quarks~\cite{Blum:2009zz}. 

There have been a few other attempts to calculate the HLO contribution from QCD, using chiral perturbation or $1/N_{c}$ expansion, or from models of QCD such as the extended Nambu-Jona-Lasinio models~\cite{deRafael:1993za} or the quark resonance model~\cite{Pallante:1994ee}. In this paper we 
use recently proposed models of QCD, called holographic QCD~\cite{Erlich:2005qh,Da Rold:2005zs,Hong:2006ta,Sakai:2004cn}, based on the gauge/gravity duality, to 
calculate the hadronic  vacuum polarization contribution. 
The holographic models of QCD were found to be quite successful to account for the properties of hadrons at  low energy and give relations to their couplings and also new sum rules~\cite{Hong:2007kx}.  
Similar attempt was made recently to estimate successfully the hadronic light-by-light contribution to the anomalous magnetic moment of muon~\cite{Hong:2009zw}.  
\section{General formulas and AdS/QCD}
\label{formular}

\begin{figure}[tbh]
\vskip 0.2in
\vskip 0.2in
	\centering
	\includegraphics[width=0.4\textwidth]{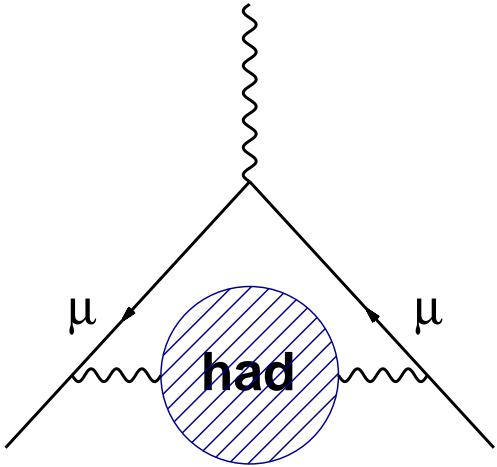}%
		\caption{\label{hlo}Leading hadronic-vacuum-polarization contribution to muon $g-2$.}
\end{figure}

Hadrons (or quarks) contribute to the anomalous magnetic moment of muon only through quantum fluctuations. The leading contribution is therefore the vacuum polarization correction, shown in Fig.~\ref{hlo},  to the internal photon line, $\Pi_{\rm em}^{\rm had}(Q^{2})$, which is defined in QCD by the electromagnetic currents of quarks, 
$J_{\rm em}^\mu = \sum_{f}\bar{q}_f \gamma^\mu Q_{\rm em} q_f$ $(f=u,d,s,\cdots)$
with the electric charge operator $Q_{\rm em}$\,, as  
\begin{equation} 
  i \int d^4 x\, e^{i qx} \langle 0 | T J_{\rm em}^\mu(x) J_{\rm em}^\nu(0)    |0\rangle 
=  ( q^2 \eta^{\mu\nu} - q^\mu q^\nu ) 
\Pi_{\rm em}^{\rm had}(-q^2) 
\,,  
\end{equation}
where $\eta^{\mu\nu}={\rm diag}(1,-1,-1,-1)$.
Since the (bare) vacuum polarization function is a divergent quantity, one should regularize it with a regulator and renormalize it by adding  a counterterm. 
Furthermore, if one uses the physical charge in the calculation of muon $g-2$, 
the renormalized vacuum function should be modified as 
\begin{equation}
\bar\Pi_{\rm em}^{\rm had}(Q^{2})=\Pi_{\rm em}^{\rm had}(Q^{2}) -\Pi_{\rm em}^{\rm had}(0) 
\end{equation}
so that the pole residue of the photon propagator does not get shifted.  

After the vacuum polarization function is obtained, we can calculate the hadronic leading order correction to the anomalous magnetic moment of muon, using  
the formula derived in~\cite{Blum:2002ii,Aubin:2006xv}~\footnote{
An alternative form of this formula, which exactly gives the same amplitude, 
is presented in Ref.~\cite{Aoyama:2008gy} }
\begin{equation} 
  a_\mu^{\rm HLO} 
  =4 \pi^2 \left(\frac{\alpha}{\pi}\right)^2 \int_0^\infty 
  d Q^2 f(Q^2) \,  \bar{\Pi}_{\rm em}^{\rm had} (Q^2) 
  \,,  \label{g-2}
\end{equation}
where $Q^2$ is the Euclidean momentum-squared and the kernel is given as  
\begin{equation}
 f(Q^2) =
  \frac{m_\mu^2 Q^2 Z^3 (1-Q^2 Z)}{1 + m_\mu^2 Q^2 Z^2}
\quad{\rm with}\;\;
Z =
- \frac{Q^2 - \sqrt{Q^4 + 4 m_\mu^2 Q^2}}{2 m_\mu^2 Q^2} \,.
\end{equation}

It is straightforward  in holographic QCD to calculate the correlation functions of QCD flavor currents. According to the gauge/gravity duality, the classical gravity action of holographic dual becomes the generating functional for the one-particle irreducible (1PI) functions of the boundary gauge theory in the large $N_{c}$ limit,  and   the bulk fields, evaluated at the UV boundary, serve as sources for the currents. 
Though the exact duality has not been established for QCD yet, as a simple model, we consider a holographic model of QCD, proposed by Erlich  et al.~\cite{Erlich:2005qh} for the calculation of the muon anomalous magnetic moment. Our analysis, however, can be easily applied to other holographic models such as Sakai-Sugimoto model~\cite{Sakai:2004cn}~\footnote{The leading hadronic contribution to muon $g-2$ has been calculated in a D3-D7 setup~\cite{Patino:2009uq}, where, however, scalar quarks are  not decoupled.}

The AdS/QCD action~\cite{Erlich:2005qh}, relevant to our present computation, is 
given in a slice of five dimensional anti de Sitter (AdS) space of unit radius as
\begin{equation} 
  S_{\rm AdS/QCD} [L,R,X] 
  =  
\int d^4 x \int_{\epsilon}^{z_m} dz \, 
  \sqrt{g}\,
{\rm Tr}\left[  - \frac{1}{4 g_5^2} (F_R^2 + F_L^2) 
+ |DX|^2 + 3 |X|^2 
\right]  
\,, \label{action}
\end{equation} 
where $g={\rm det}[g_{MN}]$, $g_{MN}=(1/z)^2 \eta_{MN}$ with $\eta_{MN}={\rm diag}(1,-1,-1,-1,-1)$, 
$F_{R,L}\equiv \partial_M R(L)_N - \partial_N R(L)_M - i [R(L)_M, R(L)_N]$ with 
the gauge fields $L_M, R_M$ of the ${\rm U}(N_f)_L \times {\rm U}(N_f)_R$ gauged-flavor symmetry in  the 5D bulk 
and $X$ denotes the bulk scalar fields,  transforming under the gauge symmetry 
as $X \to U_L X U_R^\dag$. 
One could add higher dimensional operators in the AdS/QCD action (\ref{action}) such as $|X|^{2}\,F^{2}$ or $F^{4}$, but they are suppressed by $\alpha^{\prime}$ and thus negligible at low energy. The bulk scalar gets a vacuum expectation value to break the bulk gauge symmetry down to its vectorial subgroup ${\rm U}(N_{f})$,
\begin{equation}
\left\langle X\right\rangle=\frac12  {M}\,z+\frac12{\Sigma}\,z^{3}
\end{equation}
where ${M}={\rm diag}(m_u, m_d, m_s,\cdots)$ is the current quark mass matrix and
$\Sigma$ is the chiral condensate.

\begin{figure}[tbh]
\vskip 0.2in
	\centering
	\includegraphics[width=0.25\textwidth]{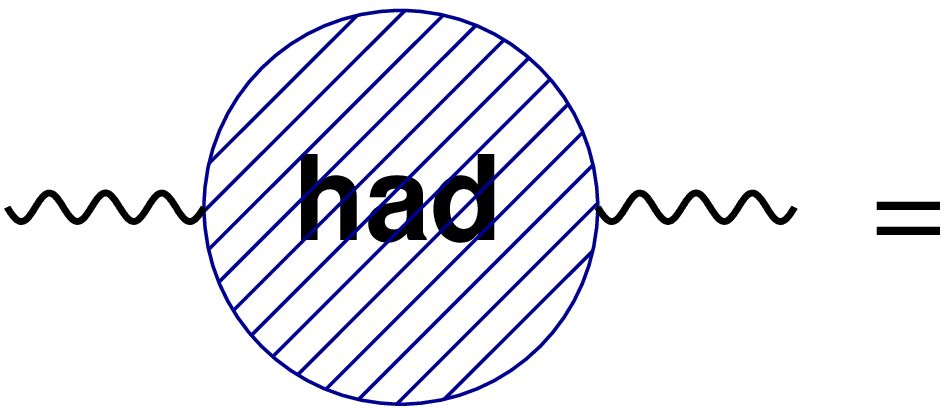}
	\includegraphics[width=0.42\textwidth]{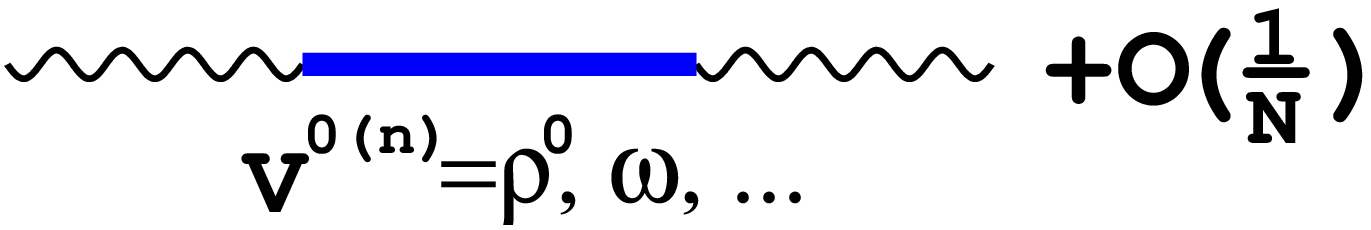}%
		\caption{\label{pivfig}A diagram illustrating neutral vector meson exchanges giving dominant contributions to $\Pi_V$ at the large $N_c$ limit.}
\end{figure}

Introducing the vector and axial vector gauge fields, 
$V=(L-R)/{2}$ and $A=(L+R)/{2}$, respectively, and working in $V_z = A_z=0$ gauge, 
we can derive the equations of motion for $V_\mu$ and $A_\mu$ 
and solve them by imposing the boundary conditions for $V_\mu$ and $A_\mu$ 
as well as for the bulk scalar field $X$, following the familiar AdS/CFT dictionary. 
One should note, however,  that 
as far as the large-$N_c$ contributions (equivalently the meson exchanges at tree-level) 
to the vector-current correlator $\Pi_V(Q^2)$ are concerned, 
we are allowed to focus only on (infinite tower of) neutral vector mesons coupled to the vector currents 
as illustrated in Fig.~\ref{pivfig}. The values of vector meson decay constants and their masses are calculable in holographic QCD.

Since  the electromagnetic charge $Q_{\rm em}=\frac{1}{3} T_0 + T_3$ in the case of two flavors, where  
the generators of flavor symmetry
$T_0=\frac{1}{2} {\bf 1}_{2 \times 2}$ and $T_3={\rm diag}(\frac{1}{2},-\frac{1}{2})$\,, 
the quark electromagnetic currents are given as 
\begin{equation} 
J_{\rm em}^{N_f=2}=\frac{1}{3} J_0 + J_3
\,.  
\end{equation} 
In the limit of exact flavor symmetry  
the vector-current correlator  is given as
\begin{equation} 
  i \int d^4 x e^{i qx} \langle 0 | T J_V^{a \mu}(x) J_V^{b \nu}(0)    |0\rangle 
= \delta^{ab} ( q^2 \eta^{\mu\nu} - q^\mu q^\nu ) 
\Pi_V(-q^2) 
\,, \label{current:V}
\end{equation} 
with $a,b$ being indices of the ${\rm U}(2)$ flavor symmetry. 
We  have then
\begin{equation} 
\Pi_{\rm em}^{N_f=2}(-q^2) 
= \frac{10}{9} \Pi_V(-q^2) 
\,. \label{piv:relate:nf2}
\end{equation} 
For $N_f=3$ we have 
$ J_{\rm em}^{N_f=3} = J_3 + \frac{1}{\sqrt{3}} J_8$ and 
$\Pi_{\rm em}^{N_f=3}(-q^2) 
= \frac{4}{3} \Pi_V(-q^2)$ in the exact limit of the ${\rm U}(3)$  flavor symmetry. 

\section{Holographic Vacuum Polarization}
Since we are interested in two-point functions of flavor currents, we need to keep terms quadratic in the bulk gauge fields for a given vacuum solution for $\left<X\right>$ in the action~(\ref{action}).  If we neglect the higher order terms like $|X|^{2}F^{2}$, the vector gauge fields do not couple to the chiral condensate $\Sigma$ but  may couple to the current quark mass operator, $M$. However, at the leading order, the neutral vectors do not couple to $X$ at all and have same mass among them, since  $[V_\mu^{0}, \left\langle X\right\rangle]=0$. 

The equations for vectors are  given at the leading order as 
\begin{equation} 
  \left( 
   \partial_z \frac{1}{z} \partial_z  + \frac{q^2}{z} 
  \right) V_{\perp \mu}^a(q,z) = 0 
  \,, \label{eq:V}
\end{equation}
where $V_{\perp \mu}^a(q,z)=v_\mu^a(q) V_\perp(q,z)$ is the transverse component of the Fourier 
transform of the vector field $V_\mu(x,z)= V_\mu^a(x,z) T^a$. 
Imposing the boundary condition 
$\partial_z V^a_\mu(q,z_m)=0$ and $V^a_\mu(q,\epsilon)=v_\mu^a(q)$, 
one gets the solution to Eq.~(\ref{eq:V}) given by Bessel functions. 
Putting the solution into the action, we are left with the boundary 
action~\cite{Erlich:2005qh}, 
\begin{equation} 
 S_{\rm eff}[v_\mu] 
 = - \frac{1}{2} \int \frac{d^4 q}{(2\pi)^4} 
v_{\mu}^a(-q) \left[ \frac{1}{g_5^2}
\frac{V_\perp(q,z) \partial_z V_\perp(q,z)}{z} \right]_{z = \epsilon} 
v^{\mu a}(q) 
 \,, \label{Seff}
\end{equation}  
which becomes 
the generating functional for the 1PI correlation  functions of the QCD flavor currents,  
 $ W^{\rm QCD}_{\rm 4 D}[v_\mu]  \equiv S_{\rm eff}[v_\mu]$\,.  

 From Eq.~(\ref{current:V}) 
we immediately read off the transverse part of the 
vector current two-point functions, 
\begin{equation} 
\frac{\delta^2 S_{\rm eff}[v_\mu]}{\delta v^{\mu a}(-q) \delta v^{\nu b}(q)}
  =   i \int d^4 x e^{i qx} \langle 0 | T J_V^{a \mu}(x) J_V^{b \nu}(0)    |0\rangle_{\perp} 
\,. \label{two-point}
\end{equation}  
Using Eq.~(\ref{two-point}) and the action~(\ref{Seff}), 
we find the vector current correlator $\Pi_V(q^2)$ in terms of the five dimensional gravity dual, 
\begin{eqnarray} 
  \Pi_V(q^2) 
  &=&   - \frac{1}{g_5^2}  \left[ \frac{\partial_z V_\perp(q,z)}{q^2\,z} \right]_{z=\epsilon} 
  \nonumber \\ 
  &=&  - \frac{1}{g_5^2} \frac{1}{q\epsilon}  
  \frac{J_0(qz_m) Y_0(q \epsilon) - Y_0(q z_m) J_0(q \epsilon)}{J_0(qz_m) Y_1(q \epsilon) - Y_0(q z_m) J_1(q\epsilon)}
\,. \label{piv:inf}
\end{eqnarray} 
Note that this $\Pi_V(q^2)$ behaves like $\ln \epsilon$ as the UV cutoff $\epsilon \to 0$, so it needs to be renormalized.  
To make this point clearer, 
we may expand Eq.~(\ref{piv:inf}) perturbatively in power of $\epsilon$ 
to obtain the asymptotic form of $\Pi_V(q^2)$,  
\begin{equation} 
  \Pi_V(q^2) 
  = \frac{1}{g_5^2} \left[
 - \frac{\pi}{2} \frac{Y_0(q z_m)}{J_0(q z_m)} 
 + \gamma - \ln 2 + \ln q \epsilon 
 + {\cal O}(\epsilon^2) 
\right]
\,, \label{piv:split}
\end{equation}
where $\gamma$ is the Euler constant. 
In order to renormalize the vacuum polarization function, 
we shall adopt a procedure of the holographic renormalization~\cite{Bianchi:2001kw,Skenderis:2002wp}, 
which is appropriate in our holographic approach.  
The holographic renormalization can be performed by introducing a counterterm 
at the UV boundary, respecting all the residual 4d-gauge symmetry, which subtracts 
only the divergent term in the vacuum polarization~(\ref{piv:split}), just like  the familiar minimal subtraction scheme. 
We find the counterterm appropriate to renormalize $\Pi_V(q^2)$,  
\begin{equation}  
  S_{\rm c}(\mu) 
  = \int d^4x \left( \frac{1}{2g_5^2} \ln \epsilon \mu   \right) 
  {\rm tr}\,[F_{\mu\nu}(x,\epsilon)]^2 
  \,, 
\end{equation}
where $\mu$ is a renormalization scale and $F_{\mu\nu}
\equiv \partial_\mu V_\nu
- \partial_\nu V_\mu
- i [V_\mu, V_\nu]$. 
This counterterm contributes to the vacuum polarization function 
\begin{equation} 
  \Pi_V^{\rm c}(q^2) 
  =  -  \frac{1}{g_5^2} \ln \mu \epsilon
  \,. \label{counterterm}
\end{equation}
Adding this counterterm contribution to Eq.~(\ref{piv:split}), 
we obtain the renormalized expression for the vacuum polarization function,
\begin{equation} 
  \Pi_V^{\rm ren}(q^2) 
\equiv 
\lim_{\epsilon \to 0}[ \Pi_V(q^2) + \Pi_V^{\rm c}(q^2)  ]
  = \frac{1}{g_5^2} \left[
 - \frac{\pi}{2} \frac{Y_0(q z_m)}{J_0(q z_m)} 
 + \gamma -  \ln 2 + \ln \frac{q}{\mu}  
\right]
\,. \label{piv:reno}
\end{equation}

We may derive another renormalized expression for $\Pi_V$ alternative to 
Eq.~(\ref{piv:reno}) by expanding Eq.~(\ref{piv:split}) in terms of 
an infinite tower of vector meson exchanges. 
This is possible due to the existence of poles in time-like momentum region 
provided by zeros of the denominator of $\Pi_V(q^2)$ of Eq.~(\ref{piv:inf}) as $J_0(q z_m)=0$. 
Let us first consider a normalizable solution $\psi_{V_n}$ 
to Eq.~(\ref{eq:V}) with $q^2$ replaced by $M_{V_n}^2$, satisfying 
the boundary condition, $\partial_z \psi_{V_n}(z_m)=0$, $\psi_{V_n}(\epsilon)=0$,  
normalized by $\int_\epsilon^{z_m} \frac{dz}{z} (\psi_{V_n}(z))^2=1$. 
The vector current correlator $\Pi_V(q^2)$ 
can then be expressed in terms of the infinite set of $\psi_{V_n}$ along with 
their poles to be with $\epsilon\to0$ 
\begin{equation} 
   \Pi_V(q^2) =  \frac{1}{g_5^2}
\sum_{n=1}^\infty \frac{[\dot{\psi}_{V_n}(\epsilon)/\epsilon]^2}{(q^2 -  M_{V_n}^2) M_{V_n}^2} 
   \,,  \label{piv:res}
\end{equation}
where the dot denotes the derivative with respect to $z$, $\dot{\psi}\equiv\partial_{z}\psi$.
From this, we read off the decay constants of the vector mesons, $F_{V_n}$, 
defined by $\langle 0|  J_V^{\mu a}(0) | V_n^b \rangle = F_{V_n} \delta^{ab} \epsilon^\mu
$ with the polarization vector $\epsilon^\mu$, 
\begin{equation} 
  F_{V_n} = \sqrt{\frac{1}{g_5^2}} \frac{\dot{\psi}_{V_n}(\epsilon)}{\epsilon}\Bigg|_{\epsilon \to 0} 
\,. \label{FVn}
\end{equation}
One can easily show that, in the expression of Eq.~(\ref{piv:res}), 
the UV divergence of Eq.~(\ref{piv:split}) is now embedded in $\Pi_V(0)$ as 
\begin{equation} 
   \Pi_V(0) =  - \sum_{n=1}^\infty \frac{F_{V_n}^2}{M_{V_n}^4}  
   =  - \frac{1}{g_5^2} \ln\frac{z_m}{\epsilon} \Bigg|_{\epsilon \to 0} 
   \,. \label{piv0}
\end{equation}
Decomposing $\Pi_V(q^2)$ as $\Pi_V(q^2)=\Pi_V(0) + [\Pi_V(q^2)-\Pi_V(0)]$ 
and replacing the first $\Pi_V(0)$ with the logarithmic divergent term of Eq.~(\ref{piv0}), 
adding the counterterm contribution of Eq.~(\ref{counterterm}) to this $\Pi_V(q^2)$, 
we arrive at another form of the renormalized 
$\Pi_V(q^2)$, $\Pi_V^{\rm ren}(q^2)$, 
\begin{eqnarray} 
   \Pi_V^{\rm ren}(q^2) 
&=&  \lim_{\epsilon \to 0} [\Pi_V(q^2) + \Pi_V^{\rm c}(q^2)] 
= - \frac{1}{g_5^2} \ln z_m \mu  
   + \sum_{n=1}^\infty \frac{q^2 F_{V_n}^2}{(q^2 - M_{V_n}^2) M_{V_n}^4} \Bigg|_{\epsilon =0} 
   \,, \label{piv:res:ren}
\end{eqnarray}
in which $(F_{V_n}/M_{V_n}^2)|_{\epsilon =0}$ is finite: 
\begin{eqnarray} 
  \frac{F_{V_n}}{M_{V_n}^2} \Bigg|_{\epsilon =0}
  &=& \sqrt{\frac{1}{g_5^2}} \frac{\sqrt{2}}{z_m M_{V_n} J_1(z_m M_{V_n})} 
\,, \label{ratio}
\end{eqnarray} 
where we have used Eq.~(\ref{FVn}) and 
\begin{equation} 
 \psi_{V_n} (z) 
 = \frac{\sqrt{2} z J_1(z M_{V_n})}{z_m J_1(z_m M_{V_n})}
\,. \label{psiVn}
\end{equation}

\section{Holographic estimation of $a_\mu^{\rm HLO}$}

In this section we calculate the leading order hadronic (HLO) contribution to  
the muon anomalous magnetic moment $a_\mu^{\rm HLO}$ in holographic QCD.

\begin{figure}[tbh]
\begin{center} 
	\includegraphics[scale=0.5]{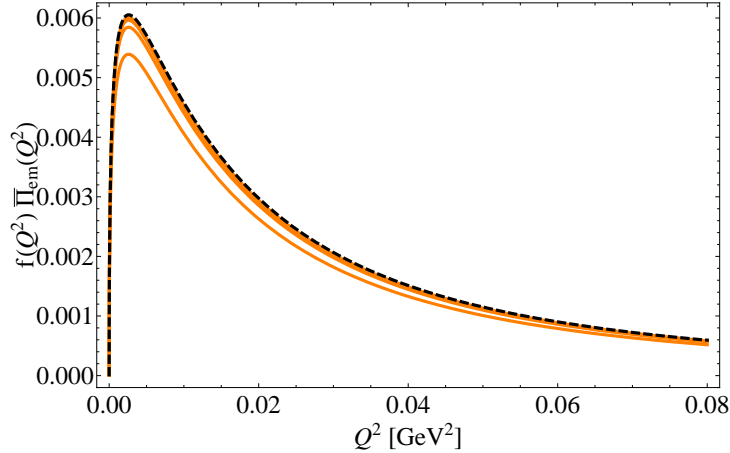}
\caption{\label{saturation}Comparison of the integral kernel 
$f(Q^2) \bar{\Pi}_{\rm em}(Q^2)$ which has a peak around $Q^2 = m_\mu^2$: 
The dashed curve corresponds to 
the result including full contributions from the infinite tower of vector mesons, 
while four bold curves are obtained by integrating out the infinite tower 
at the levels of $n=1,2,3,4$. 
The dashed curve is almost (within about 1\% deviation) reproduced when $n=4$.  
In the plot the number of flavors $N_f$ is taken to be 2.} 
\end{center} 
\end{figure}

We should first note that the holographic description of the present model breaks down 
at a high-energy scale above which the underlying stringy ingredients would not be 
negligible. 
Recall, furthermore, that the vector current correlator $\Pi_V$ includes 
the UV divergence which has been renormalized and converted into 
the renormalization scale based on the holographic renormalization scheme. 
This implies that the HLO calculation of $\Pi_V$ is potentially sensitive to 
the UV scale/renormalization scheme. 
 From these points of view, 
 it seems reasonable to 
truncate the infinite tower of vector mesons at a finite level $n$. 
The simplest way to do  is to use Eq.~(\ref{piv:res:ren}) and 
expand it in powers of $(q^2/M_{V_n}^2)$, and keeping only a few excited states. 
Examining 
the effects from higher resonances on the integral kernel of Eq.~(\ref{g-2}) [See Fig.~\ref{saturation}], 
we find that truncating the tower at $n=4$ suffices within about 1\% accuracy 
to saturate the full contributions. 
We therefore truncate at the 4-th ($n=4$) excited states and derive an expression of ${\Pi}^{\rm ren}_V(q^2)$ valid 
for a low-energy region $M_{V_1}^2 \lesssim q^2 \ll \{ M_{V_5}^2, \cdots \}$, 
to get
 \begin{eqnarray} 
  \Pi^{\rm ren}_V(q^2)
  &=& 
- \frac{1}{g_5^2} \ln z_m \mu 
+  \sum_{n=1}^4 \frac{q^2 F_{V_n}^2}{(q^2 - M_{V_n}^2)M_{V_n}^4} 
+   {\cal O}(q^2/(M_{V_5}^2))
  \, . \label{piv:int}
\end{eqnarray}

In order to utilize the formula Eq.~(\ref{g-2}), we subtract 
$\Pi_V^{\rm ren}(0)$  to obtain 
\begin{eqnarray} 
  \bar{\Pi}_V(Q^2) 
  &\equiv & \Pi_V^{\rm ren}(Q^2) - \Pi_V^{\rm ren}(0) 
  \,, \nonumber \\ 
  &=& 
  \sum_{n=1}^4 \frac{Q^2 F_{V_n}^2}{(Q^2 + M_{V_n}^2)M_{V_n}^4} 
  +   {\cal O}(Q^2/(M_{V_5}^2))
  \, , \label{bar:pi}
\end{eqnarray} 
where $Q^2=-q^2$ is Euclidean momentum-squared. 
Note that the dependence of the renormalization scale $\mu$ has been 
removed.

 Let us now evaluate $\bar{\Pi}_V$ of Eq.~(\ref{bar:pi}) numerically. 
Looking at Eq.~(\ref{bar:pi}), to this aim, we see that all we need to do is 
calculate the values of the vector meson poles $M_{V_n}$ and their pole residues 
$(F_{V_n}/M_{V_n}^2)$ for each $n=1,2,3,4$. 
We first notice that the mass hierarchy of vector mesons are completely determined by 
locations of zeros given as $J_0(z_m M_{V_n})=0$ as seen from Eq.~(\ref{piv:inf}). 
Especially, for $n=1,2,3,4$, we have 
\begin{eqnarray} 
&& 
z_m M_{V_1} = 2.405 
\,, \qquad 
z_m M_{V_2} = 5.520
\,,  \nonumber \\ 
&&  z_m M_{V_3} = 8.654  
\,, \qquad 
  z_m M_{V_4} = 11.792 
\,. 
\end{eqnarray}
Therefore, once the lowest pole is identified with the $\rho$ meson pole, i.e., 
$M_{V_1}=M_\rho=0.775 \, [{\rm GeV}]$~\cite{Amsler:2008zz} which fixes 
$z_m=3.101 \, [{\rm GeV}]^{-1}$, 
we can obtain a set of the vector meson masses (in unit of GeV);
\begin{eqnarray} 
&&
M_{V_1} = 0.775 \quad ({\rm fit})  
\,, \qquad 
M_{V_2} = 1.779 \,, 
  \nonumber \\ 
&& 
M_{V_3} = 2.788  
\,, \hspace{55pt}
M_{V_4} = 3.802  
\,, \label{rho-masses}
\end{eqnarray} 
in which the values of $M_{V_2},M_{V_3}$ are compared to the empirical ones~\cite{Amsler:2008zz}, 
$M_{\rho_2}|_{\rm exp}=1.45$ and $M_{\rho_3}|_{\rm exp}=1.70$ in unit of GeV. 
We next turn to calculating the pole residues $F_{V_n}/M_{V_n}^2$ expressed as in 
Eq.~(\ref{ratio}). 
Note that the overall factor $(1/g_5^2)$ of Eq.~(\ref{ratio}) is determined by considering a fit with 
the expression derived from the operator product expansion of $\Pi_V(Q^2)$, 
and it is fixed as 
\begin{equation} 
  \frac{1}{g_5^2} 
  = \frac{N_c}{12 \pi^2}
\,. 
\end{equation}
  For $n=1,2,3,4$, 
the values of $F_{V_n}/M_{V_n}^2$ of Eq.~(\ref{ratio}) 
are then calculated up to the overall factor as 
\begin{eqnarray} 
&&    
\frac{F_{V_1}}{M_{V_1}^2} \cdot (\sqrt{N_c/12\pi^2})^{-1}  
=
1.133
\,, \qquad 
\frac{F_{V_2}}{M_{V_2}^2}\cdot (\sqrt{N_c/12\pi^2})^{-1}   
= 
-0.753\,,
\nonumber\\ 
&& 
\frac{F_{V_3}}{M_{V_3}^2} \cdot (\sqrt{N_c/12\pi^2})^{-1}  
= 
0.602 
\,, \qquad 
 \frac{F_{V_4}}{M_{V_4}^2} \cdot (\sqrt{N_c/12\pi^2})^{-1}  
=
- 0.516  
\,. \label{FVratios}
\end{eqnarray}

Now we are ready to evaluate the HLO contribution to the muon 
anomalous magnetic moment $a_\mu^{\rm HLO}$. 
We shall focus on a two-flavor ($N_f=2$) HLO contribution, 
coming from the lightest two flavors $(u,d)$, 
and then compare the predicted value of $\alpha_\mu^{\rm HLO}$ 
with that estimated from the new BABAR data on the $e^+ e^- \to \pi^+\pi^-$ 
events detected in a range of center of mass energies 
$2 m_\pi \le \sqrt{s} \le 1.8$ GeV~\cite{Aubert:2009fg}. 
Taking into account Eq.~(\ref{piv:relate:nf2}) and substituting $\bar{\Pi}_V$ of Eq.~(\ref{bar:pi}) 
into the formula Eq.~(\ref{g-2}), 
using the theoretical values for the vector meson poles and their pole residues 
listed in Eqs.(\ref{rho-masses}) and (\ref{FVratios}), 
we obtain 
\begin{equation} 
  a_\mu^{\rm HLO}|^{N_f=2}_{\rm AdS/QCD} 
=470.5 \times 10^{-10}, 
\label{U2:result}
\end{equation} 
which agrees, within 30\% errors, 
with the currently updated value~\cite{Aubert:2009fg} 
\begin{equation} 
 a_\mu^{\rm HLO}[\pi\pi] |_{\rm BABAR} 
=(514.1 \pm 3.8 )\times 10^{-10}
\,. \label{BABAR}
\end{equation}
We expect that the $1/N_c$ corrections together with the isospin-breaking corrections could make 
up for the discrepancy between the values of Eq.~(\ref{U2:result}) and Eq.~(\ref{BABAR}).

\section{Discussion} 

We have calculated the hadronic leading correction  (HLO) to the muon magnetic anomaly with two light flavors. 
The result is compared with a recently updated HLO value~\cite{Davier:2009zi}, 
estimated by including not only the BABAR new $e^+ e^- \to \pi^+ \pi^-$ data~\cite{Aubert:2009fg} 
but also the other experimental data involving other inclusive decay processes 
in which multi-hadrons more than 
two $\pi$s, such as $2\pi^0 \pi^+\pi^-$, $2\pi^+ 2 \pi^-$, are included in 
the final states, as well as some exclusive decay processes.  
 From a theoretical point of view, 
those multi-hadronic decaying processes are thought to be 
dominated by higher order contributions of meson-loops 
in the large $N_c$ expansion. However, those higher order corrections  
have not been achieved in the present holographic framework which is based 
on the holographic correspondence, established only at the large $N_c$ limit. 
It is not adequate, therefore, to compare the result obtained from 
our present holographic calculation with that deduced from data involving 
such those decaying processes. 

Similarly,  we have not compared with the value estimated 
based on lattice calculation~\cite{Blum:2002ii} with the staggered fermions of $N_f=3$,  
because the value of Ref.~\cite{Blum:2002ii} includes a part of the meson-loop contributions. 

We suspect that incorporation of $1/N_c$ corrections as done in Ref.~\cite{Harada:2006di} 
might make it possible to compare with these references.

\subsection{Acknowledgements}
We thank M. Drees and M. Hayakawa for useful comments and informing us of the references~\cite{Aubert:2009fg,Davier:2009zi} and~\cite{Aoyama:2008gy}, respectively. 
The work of D.~K.~H. and S.~M. was supported by the Korea Research
Foundation Grant funded by the Korean Government (KRF-2008-341-C00008) and D.~Kim was supported by  the KRF grant (KRF-2008-313-C00162).

\end{document}